\documentstyle[aps,prl,psfig]{revtex}
\begin{document}
\wideabs{
\title{The joys and pitfalls of Fermi surface mapping in Bi$_2$Sr$_2$CaCu$_2$O$_{8-\delta}$ using angle resolved photoemission.}
\author{S. V. Borisenko$^1$, M. S. Golden$^1$, S. Legner$^1$, T. Pichler$^1$, C.D\"urr$^1$, M.Knupfer$^1$, J. Fink$^1$, G.Yang$^2$, S. Abell$^2$, H.Berger$^3$}

\address{$^1$ Institut f\"ur Festk\"orper- und Werkstofforschung Dresden, P.O.Box 270016, D-01171 Dresden, Germany}
\address{$^2$ School of Metallurgy and Materials, The University of Birmingham, Birminhgam, B15 2TT, United Kingdom}
\address{$^3$ Institut de Physique Appliqu\'ee, Ecole Politechnique F\'ederale de Lausanne, CH-1015 Lausanne, Switzerland}
%\date{\today}
\maketitle
\begin{abstract}

On the basis of angle-scanned photoemission data recorded using {\it unpolarised} radiation, with high (E,{\bf k}) resolution, and an extremely dense sampling of {\bf k}-space, we resolve the current controversy regarding the normal state Fermi surface (FS) in Bi$_2$Sr$_2$CaCu$_2$O$_{8-\delta}$ (Bi2212).
The true picture is simple, self-consistent and robust: the FS is hole-like, with the form of rounded tubes centred on the corners of the Brillouin zone. Two further types of features are also clearly observed: shadow FSs, which are most likely to be due to short range antiferromagnetic spin correlations, and diffraction replicas of the main FS caused by passage of the photoelectrons through the modulated Bi-O planes.
\end{abstract}
}
The topology and character of the normal state Fermi surfaces of the high temperature superconductors have been the object of both intensive study and equally lively debate for almost a decade.
Angle-resolved photoemission spectroscopy (ARPES) has played a defining role in this discussion.
In particular, the pioneering work of Aebi {\it et al.} illustrated that angle-scanned photoemission using unpolarised radiation can deliver a direct, unbiased image of the complete FS of Bi2212 \cite{Aebi}, confirming the large FS centered at the corners of the Brillouin zone predicted by band structure calculations \cite{Calculations}.
Furthermore, the use of the mapping method enabled the indentification of weak additional features (dubbed the shadow Fermi surface or SFS) which were attributed to the effects of short-range antiferromagnetic spin correlations, the existence of which had already been proposed theoretically \cite{Kampf}.
Subsequently, conventional ARPES investigations (involving the analysis of series of energy distribution curves (EDCs) along a particular line in k-space) clearly identified a further set of dispersive photoemission structures which are extrinsic and result from a diffraction of the outgoing photoelectrons as they pass through the structurally modulated Bi-O layer, which forms the cleavage surface in these systems \cite{Ding}.

Recently, this whole picture of the normal state FS of Bi2212 has been called into question. 
ARPES data recorded using particular photon energies (32-33 eV) have been interpreted in terms of either: a FS with missing segments \cite{Saini1}, an extra set of one dimensional states \cite{Saini2}, or an electron-like FS centred around the $\Gamma$ point \cite{Chuang}.
A further study suggests that both electron or hole-like FS pieces can be observed merely depending on the photon energy used in the ARPES experiment \cite{Feng}.

These points illustrate that the situation as regards the true topology and character of the normal state FS of Bi2212 as seen by photoemission spectroscopy is, in fact, far from clear.
Thus, considering both the fundamental significance of the FS question in general, and the importance of photoemission in this debate, it is essential that an unambiguous framework is arrived at for the interpretation of the ARPES data.

In this Letter we present angle-scanned photoemission data from pure and Pb-doped Bi2212.
As our data-sets contain more than 1000 high (E,{\bf k})-resolved EDCs per Brillouin zone quadrant, we combine the advantages of both the mapping and EDC methods, making data manipulation such as interpolation superfluous.
We show that the origin of the recent controversy stems from the simultaneous presence of three different types of photoemission features around the $\overline{\text{M}}$ point \cite{XYM}: the main FS, diffraction replicas (DRs) and the SFS.

The ARPES experiments were performed using monochromated, unpolarised He I radiation and a SCIENTA SES200 analyser enabling simultaneous analysis of both the E and {\bf k}-distribution of the photoelectrons. 
The overall energy resolution was set to 30 meV and the angular resolution to $\pm$0.38 $^\circ$, which gives $\Delta${\bf k} $\leq$ 0.028 \AA$^{-1}$ (i.e. 2.4 $\%$ of $\Gamma$X).
High quality single crystals of pristine \cite{Bi2212crystals} and Pb-doped Bi2212, the latter grown from the flux in the standard manner, were cleaved in-situ to give mirror-like surfaces and all data were measured at either 120 or 300K within 6 hours of cleavage.

Figures 1a and 1b show series of photoemission data (T$\sim$ 300K) of Bi2212 taken along the two high symmetry directions $\Gamma$X and $\Gamma$Y in {\bf k}-space\cite{XYM}, respectively.
The EDCs (right panels) are shown together with 2D (E, {\bf k}) 
representations where the photoemission intensity\cite{Normalization} forms the grey scale (left panels).

We deal first with $\Gamma$X. 
Starting from the $\Gamma$ point, we clearly observe the main CuO$_2$ derived band crossing the Fermi level, $E_{\text F}$, at $\sim$ 0.4($\pi$,-$\pi$). 
Also evident are two weaker features straddling the X point, which cross $E_{\text F}$

\begin{figure} 
\begin{center}
\leavevmode \psfig{file=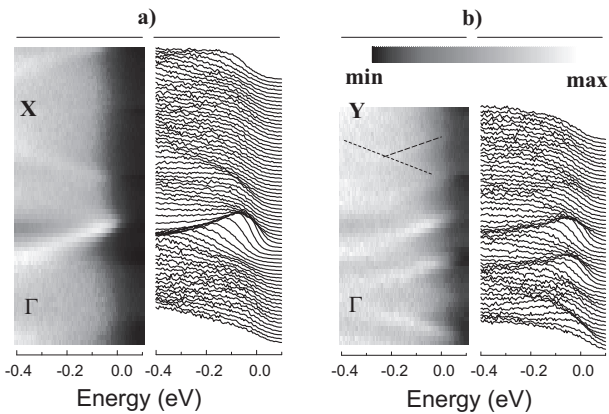,width=8cm,angle=0}
\end{center}
\caption{ARPES of Bi2212 recorded at 300K along (a) $\Gamma$X  and (b) $\Gamma$Y. In each case, the right panels show the EDCs and the left panels show (E,{\bf k})-plots in which the photoemission intensity is represented by the grey scale.}
\label{EDCs}
\end{figure}

at wavevectors equivalent to the main band crossings but shifted by the vector ($\pi$, -$\pi$).
These are the shadow bands first observed in Ref. \onlinecite{Aebi}.
A strikingly different picture occurs for $\Gamma$Y (parallel to the crystallographic b-axis, Fig. 1b).
Once again, the strongest feature is the main band, but now additional, extrinsic DRs of the main band are clearly visible, shifted in {\bf k} from the main band by {\it n}{\bf q}=(0.21$\pi$,0.21$\pi$), whereby {\it n} is the order of the DR \cite{Ding}.
Thus, starting from the bottom of Fig.\ref{EDCs}(b) we see firstly a first-order DR of the main band (the latter crosses $E_{\text F}$ at $\sim$0.4(-$\pi$,-$\pi$)).
Then, around the $\Gamma$ point a feature resulting from the overlap of very weak second order DRs is observed.
There then follows a 1st order DR of the parent main band which is then itself seen crossing $E_{\text F}$ at $\sim$0.4($\pi$,$\pi$)
Subsequently, a further 1st order DR of the same main band is seen, followed by indications for the shadow band and a 2nd order DR.
It should be stressed that the strongly dispersive nature of the states along  $\Gamma$X,Y eases the task of interpreting the photoemission data. 

Thus, the data shown in Fig. 1 indicate the presence of three different types of features for the 
$\Gamma$X,Y photoemission data from Bi2212 related to the main bands, the shadow bands and
diffraction replicas of the main bands.
In this point there is a broad consensus \cite{Aebi,Ding,Chuang}.

The picture for the region around the $\overline{\text{M}}$-point, however, is currently the subject of considerable controversy.
In fact, the interpretation of the photoemission data from this region of {\bf k}-space is the key to resolving this debate and settling, once and for all, the true FS nature and topology in Bi2212.

Therefore, in Fig. 2 we present a detailed momentum map of the normal state (T=120K) Fermi surface of Bi2212 around the $\overline{\text{M}}$-point.
The image is based upon 1300 EDCs \cite{Normalization}, each of energy width 500$\le$ $E_{\text B}$ $\le$ -100 meV - and thus combines the unbiased view given by a map with the security of being able to examine a full EDC at each particular {\bf k}-point.
Following the discussion of Fig. 1 we have added guidelines indicating the location of the main FS (thick black solid line), the DRs (1st order: thin black solid line (on top of the map); 2nd order: dashed black line) and the shadow features (thick red solid line).

The main FS appears in Fig. 2 as arcs of high intensity centred around the X and Y points.
There is absolutely no indication of a 'closure' of the two main FS arcs shown in Fig. 2 at (0.8$\pi$,0) as reported in Refs. \onlinecite{Chuang,Feng}.
Thus there is no $\Gamma$-centered (electron-like) FS. 
The diffraction replicas of the main FS are also clearly present, both individually (e.g. the sole (red) feature on the dark blue background along the $\Gamma$-$\overline{\text{M}}$-Z direction at (1.5$\pi$,0) is a 1st order DR) and collectively (in principle up to infinite order), leading to a bundling of intensity along a ribbon centered on the (0,-$\pi$)-($\pi$,0) line - indicated in Fig. 2 by grey shading. 

\begin{figure}
\begin{center}
\leavevmode \psfig{file=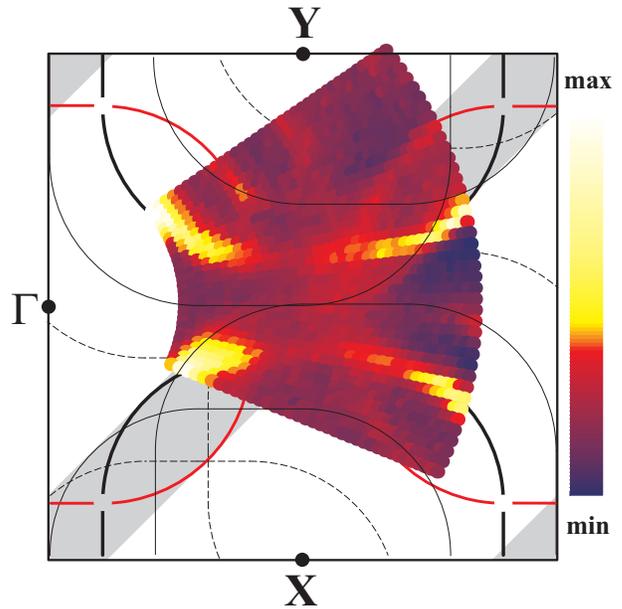,width=8cm,angle=0}
\end{center}
\caption{Normal state (T=120K) Fermi surface map of Bi2212. The colour scale indicates the photoemission intensity at $E_{\text F}$ [11].  The guidelines indicate the main FS (thick, black solid line); shadow FS (solid red line); 1st-order DRs (thin solid black line, on top of map); 2nd order DRs (dashed black line). For details see text.}
\label{LNmap}
\end{figure}

The intensity distribution in the ribbon is, however, anisotropic.
The edges of the ribbon are more intense than the centre, reflecting the intensity distribution around the main FS (the DR's will necessarily have the same intensity distribution as the main FS).
Finally, we point out the SFS, which follows the red lines underlying the map and is seen in these data more clearly than ever before. 

The main point that is clear from Fig. 2 is the richness of structure in the ARPES data around $\overline{\text{M}}$. This arises from the complex interplay between main FS, DRs and the SFS features. As an example of this, one can see the overlap of the SFS and 1st order DR features at ca. 0.6($\pi$,$\pi$) as a bright spot on the map.

We emphasize that only an analysis of uninterpolated data recorded with high (E,{\bf k})-resolution on an extremely fine {\bf k}-mesh can enable the discrimination between the numerous features concentrated within this small region of the Brillouin zone.

As a test of the robustness of the picture developed above, we show in Fig. 3 a large Fermi surface map of Pb-doped Bi2212.
This time 3760 EDCs form the basis of the image. 
We chose the Pb-doped material as it does not possess the strong structural modulation along the crystallographic b-direction which is characteristic of Bi2212\cite{Aebi_PbdopedBSCCO}.
This should result, then, in the disappearance of DR-related features in the map. 

\begin{figure}
\begin{center}
\leavevmode \psfig{file=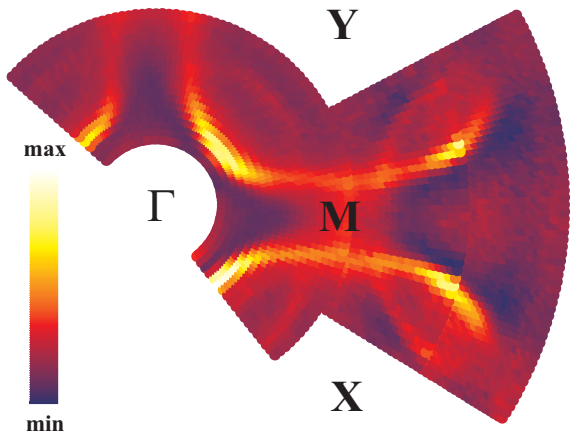,width=7.5cm,angle=0}
\end{center}
\caption{Normal state (T=120K) Fermi surface map of Pb-doped Bi2212. The colour scale indicates the photoemission intensity at $E_{\text F}$ [11]. 
The main FS has the topology of rounded tube-sections, and is hole-like, centred upon the X,Y points. The shadow FS is also clearly visible. For details see text.
}
\label{Pbmap}
\end{figure}

As can be seen from Fig. 3, this is quite evidently the case as there is no intensity ribbon centred along the (0,-$\pi$)-($\pi$,0) line and intensity profile across the map is practically symmetrical about the $\Gamma$-$\overline{\text{M}}$-Z line.
In this way we can put the assignment of the DR features in the Bi2212 data beyond any doubt.

The main FS, which is presented here with unprecedented clarity, has the form of tubes centered 
around the X,Y points.
There are absolutely no indications of a $\Gamma$-centred, electron-like FS.
The detailed topology of the main FS differs from that in the recent literature in a number of important points, which are related to the favourable experimental conditions used in this study.

Firstly, the use of unpolarised radiation reduces the distorting effects of the photoemission matrix elements to a minimum - illustrated by the roughly equal intensity of the tube sections in both the $\Gamma$X and $\Gamma$Y directions.
All of the published FS investigations other than those of Aebi and co-workers have used
highly polarised synchrotron radiation and are therefore suscecptible to extreme suppression
of intensity in particular {\bf k}-space regions due purely to symmetry-related matrix element effects.
In this way, mapping measurements using polarised radiation are inherently at a disadvantage when one wishes to determine the detailed FS topology without having to know, a priori, what it is.
For example, Fermi surface maps of Bi2212 showed a strong dependence on the experimental geometry (and thus
polarisation conditions) \cite{Saini1,Bianconi}.
Having made this point, it is obvious that maps recorded using polarised radiation cannot be used to support contentions \cite{Chuang} of an electron-like FS in Bi2212.

The second point concerns the exclusive use of {\it real} experimental data. 
If a coarse {\bf k}-mesh is used to construct a map via interpolation, artefacts can result whereby
the topology of the Fermi surface is extremely sensitive to the procedure used to define ${\bf k}_{\text F}$.
As an illustration of this, we comment on the recent FS map of Feng et al. \cite{Feng}, derived from data recorded using polarised radiation at 51 k-points in the (0,0)-($\pi$,$\pi$)-(0,$\pi$) octant of the Brillouin zone, which shows nested FS segments only when using the $\nabla$n({\bf k}) method of determining ${\bf k}_{\text F}$ \cite{Feng}. 
The FS data presented here for both pristine and Pb-doped Bi2212 show a more curved FS arc in this region of {\bf k}-space, although a degree of parallelity across the $\overline{\text{M}}$-point does exist.
The important point here is that the FS topology in ARPES data-sets of sufficient quality is extremely robust - the maps shown in Figs. 2 and 3 can be redrawn using either the $I_{\text max}$ or $\nabla$n({\bf k}) methods of determining ${\bf k}_{\text F}$, without altering the form of the main FS.

The third point in the comparative discussion of our data with that in the literature concerns
the question of photon energy. 
We see no indication for an electron-like FS in either 2212 systems studied, neither from the He I
data presented here, nor from full FS maps of Pb-Bi2212 (not shown) recorded using unpolarised He II radiation (h$\nu$=40.8eV).
We reach the same conclusion from the analysis of our synchrotron ARPES data from Bi2212 \cite{Footnote_SRConditions}, in which no sign of a main band crossing is observed along $\Gamma$$\overline{\text{M}}$ for photon energies of either 40 or 50 eV (data not shown).
If we use h$\nu$=32eV, we observe a strong suppression of the intensity of the saddle-point singularity states near $\overline{\text{M}}$.
As the Fermi surface of a material cannot depend on the photon energy used to measure it, it is clear that photon energies between 32-33 eV (as used in Refs. \onlinecite{Saini1,Saini2,Chuang,Feng,Bianconi}) are not suited to giving an undistorted picture of the true FS of Bi2212.

Indeed, our data offer a natural explanation for the observations made with h$\nu$=32-33eV. 
The matrix-element mediated supression of the saddle point intensity for these photon energies
explains both the suppression of spectral weight directly along (0,-$\pi$)-($\pi$,0) line observed in \cite{Saini1},
and means that the edges of the ribbon mentioned earlier will become relatively more intense.
Thus, traversal of the two edges of the ribbon feature could be mistakenly interpreted as a 'main'
band crossing along the $\Gamma$-$\overline{\text{M}}$ line\cite{Chuang,Feng}.

Finally, we turn to the shadow FS features. 
The most mundane explanation would be an extrinsic diffraction of the CuO$_2$-plane photoelectrons
on their way to or through the surface of the crystal, giving rise to full, tube-like copies of the main FS -
i.e. the SFSs are mere DRs.
Within this picture, as there is no evidence for reconstruction of the Sr-O or Bi-O layers of Bi2212 other than the well-known Bi-O modulation, the SFS features would then have to be {\it fifth}-order DRs (assuming {\bf q} is exactly 0.2($\pi$,$\pi$), which is not the case). 
For Bi2212 this scenario is clearly highly unrealistic, considering the relative intensities of the first and second-order DR features seen in Figs. 1 and 2.
In addition, if the SFS were a DR, it should show the same intensity distribution as the correponding sections of the main FS (as is seen for the genuine Bi2212 DRs observed due to the Bi-O-modulation).
As this is not the case for the SFSs in either pristine or Pb-doped Bi2212, it would appear that they are intrinsic to
the CuO$_2$ planes and thus due to the presence of a Brillouin zone of half the size of the original.
Such a reduction of the Brillouin zone could be due to structural effects, as has been illustrated for the case of Bi2201 \cite{Singh}.
However, there exists no evidence of this in the case of Bi2212. 

Thus, at least for pristine Bi2212, all indicators point to a spin-related origin of the shadow FS in the photoemission data, as originally proposed \cite{Aebi,Kampf}.
As there is electron diffraction evidence of a c(2x2)-like reconstruction in Pb-doped Bi2212 \cite{Aebi_PbdopedBSCCO}, no definitive conclusion is possible at this stage regarding the origin of the SFSs in this compound.
Nevertheless, the strong similarities as regards the topology and relative intensity of the shadow FSs in both the
pristine and Pb-doped systems would seem to point to a common origin for these features.

Taking the SFSs to be intrinsic to the CuO$_2$ planes, regardless their detailed origin, the overall FS topology
cannot be that of the main FS tubes with additional SFS-tubes shifted by ($\pi$,$\pi$), as this would lead to a crossing
of the two FS pieces (which is forbidden), as well as the necessity of accepting the simultaneous presence
of hole-like and electron-like FSs of the same size originating from the same CuO$_2$ planes.
Thus, the two FS pieces will be separated by a gap, whose magnitude depends on the interaction 
responsible for the Brillouin zone reduction.
This gap could, however, be small enough to remain undetected in a photoemission experiment.

In conclusion, we have shown that high (E,{\bf k}) resolution, high {\bf k}-density angle-scanned photoemission data-sets combining the advantages of both the mapping and EDC approaches give a self-consistent and robust picture of the nature and topology of the FS in the 2212-based materials.
From a comparison of pristine and Pb-doped Bi2212 we have shown there to be three different features in the ARPES data of Bi2212:

- the main FS is hole-like, with the topology of a curved tube centred around the X,Y points;

- the Bi-O modulation gives rise to extrinsic diffraction replicas of the main FS which lead to high intensity ribbons centered on the (0,-$\pi$)-($\pi$,0) line;

- a shadow FS is also clearly present, 
which at least in the case of Bi2212, is likely to be of spin-related origin.

{\it Note added:} while completing this paper, we became aware of a preprint \cite{Fretwell} containing high k-density ARPES data of Bi2212 recorded using polarised radiation. These data, although measured deep in the superconducting state, are used to confirm the hole-like nature of the main normal-state FS, and the distorting effect of matrix elements on the photoemission spectra when using h$\nu$=33eV.

We are grateful to the the BMBF (05 SB8BDA 6), the DFG (Graduiertenkolleg 'Struktur- und Korrelationseffekte in Festk\"orpern' der TU-Dresden) and the SMWK (4-7531.50-040-823-99/6) for financial support, to U. J\"annicke-R\"ossler and K. Nenkov for characterisation of the crystals.


\begin{references}
\bibitem{Normalization} Each EDC is normalised by division with the total integrated intensity of the spectrum. This gives a new criterion for determining {\bf k}$_{\text F}$, and can be viewed as a combination of the $I_{\text max}$ and $\nabla$n({\bf k}) methods.
\bibitem{Footnote_SRConditions} Spectra were recorded at 100K with $\Delta\Theta$ = $\pm$1$^\circ$ and $\Delta$E = 70 meV using linearly polarised synchrotron radiation from the U2-FSGM beamline at the BESSY I facility.
\end{references}
\end{document}